\documentclass[12pt,letterpaper]{article}
\usepackage[utf8]{inputenc}
\usepackage{amsmath, amscd}
\usepackage{amsfonts}
\usepackage{amssymb}
\usepackage{amsthm}  
\usepackage[top=1in, bottom=1in, left=1in, right=1in]{geometry}
\usepackage{titlesec}
\usepackage{hyperref}
\usepackage[]{footmisc}
\usepackage{mathrsfs}
\usepackage{graphicx}
\usepackage{verbatim}
\usepackage{algpseudocode}
\usepackage{textcomp}
\usepackage{color}
\usepackage{csquotes}
\usepackage{commath}
\usepackage{float}
\titleformat{\section}{\normalsize\scshape\center}{\thesection}{1em}{}
\titleformat{\subsection}{\normalsize\scshape\center}{\thesubsection}{1em}{}
\makeatletter
\renewcommand\@makefntext[1]{%
    \parindent 1em%
    \@thefnmark.~#1}
\makeatother

\makeatletter
\def\moverlay{\mathpalette\mov@rlay}
\def\mov@rlay#1#2{\leavevmode\vtop{%
   \baselineskip\z@skip \lineskiplimit-\supdimen
   \ialign{\hfil$\m@th#1##$\hfil\cr#2\crcr}}}
\newcommand{\charfusion}[3][\mathord]{
    #1{\ifx#1\mathop\vphantom{#2}\fi
        \mathpalette\mov@rlay{#2\cr#3}
      }
    \ifx#1\mathop\expandafter\displaylimits\fi}
\makeatother

\newtheorem{innercustomgeneric}{\customgenericname}
\providecommand{\customgenericname}{}
\newcommand{\newcustomtheorem}[2]{%
  \newenvironment{#1}[1]
  {%
   \renewcommand\customgenericname{#2}%
   \renewcommand\theinnercustomgeneric{##1}%
   \innercustomgeneric
  }
  {\endinnercustomgeneric}
}

\newcustomtheorem{ctheorem}{Theorem}
\newcustomtheorem{clemma}{Lemma}
\newcustomtheorem{cdefinition}{Definition}
\newcustomtheorem{cproposition}{Proposition}
\newcustomtheorem{ccorollary}{Corollary}
\newcustomtheorem{cexample}{Example}

\newtheorem{definition}{Definition}

\theoremstyle{remark}

\theoremstyle{plain}

\begin{document}

\title{An antidote for hawkmoths: on the prevalence of structural chaos in non-linear modeling\footnote{We would like to thank Mathias Frisch, Blaine Lawson, Seung-Yeop Lee, Connor Mayo-Wilson, Jessica Williams, the audience members at talks in London, Ontario and Chicago, and our anonymous reviewers for helpful comments on earlier drafts as well as many useful discussions of mathematical questions.  Responsibility for any remaining errors is of course our own!}}

\author{\small{\textsc{Lukas Nabergall, Alejandro Navas, and Eric Winsberg}}}
\date{\small{\textsc{\today}}}
\maketitle
% \(\)
\allowdisplaybreaks

\begin{abstract}

This paper deals with the question of whether uncertainty regarding model structure, especially in climate modeling, exhibits a kind of ``chaos.''  Do small changes in model structure, in other words, lead to large variations in ensemble predictions?   More specifically, does model error destroy forecast skill faster than the ordinary or ``classical" chaos inherent in the real-world attractor?  In some cases, the answer to this question seems to be ``yes."  But how common is this state of affairs?  Are there precise mathematical results that can help us answer this question?  And is dependence on model structure ``sensitive'' in that arbitrarily small errors can destroy forecast skill?  We examine some efforts in the literature to answer this last question in the affirmative and find them to be unconvincing.

\end{abstract}

\tableofcontents

%%%%%%%%%%%%%%%%%%%%%%%%%%%%%%%%%%%%%%
\section{Introduction}
%%%%%%%%%%%%%%%%%%%%%%%%%%%%%%%%%%%%%%

Numerical modeling plays a central role in the science of climate, and numerical climate models play a central role in producing the predictions and projections that are intended to guide policy makers in their decision making tasks.   But the perfect model of the climate is not available, and it will not be available to us anytime soon.  Thus, uncertainty in our climate predictions and projections due to uncertainty in model structure is a topic that is of central importance to climate science.  It is a topic that is especially important to the kind of climate science that is intended to guide decision making.

This paper deals with the issue of whether structural uncertainty in models is in some sense ``chaotic.''  It deals, in other words, with the question of whether small changes in model structure lead to large variations in ensemble predictions.  And it deals with the question of what the contours of that problem are: how common, and how severe, is structural chaos?

In climate science and nearby disciplines, the fact that small changes in model structure \textit{can} produce large changes in model output is not really open to debate.  But it is not the norm.  A good deal of relevant knowledge about this issue comes from the experiences of climate scientists and meteorologists in using and testing their models.   Most of the time small variations in model structure (such as the tuning of parameters) do not lead to big changes in climatology, emergent patterns, or initial condition forecasts. Empirically, modelers report the sense that the structural variations of many sorts are not `chaotic' (even while each model version is chaotic in the classical sense).\footnote{(Doblas-Reyes et al, 2013).} 

However, we do know that large perturbations are possible by adding new or different physics. A reviewer has kindly provided a few examples to illustrate the point: The addition of viscosity to the Euler equations (to get the Navier-Stokes equations)---even when in the majority of the domain the viscosity plays no role---has major implications for the flow. Airplanes could not fly without the viscosity that allows them to set up a circulation around their wings, even though lift is a purely Eulerian concept. Introducing heterogeneous chemistry on the surface of polar stratospheric clouds gives rise to wholly new phenomena in the presence of CFCs (the polar ozone hole) that no amount of initial condition variation would ever have implied. Some models predict massive collapses in the North Atlantic Meriodonal Overturning Circulation as a function of global warming, and others, with similar (though not identical) ocean mixing processes do not. Another example related to initialized forecasts is the differing predictions of the track of Hurricane Sandy by the ECMWF and NCEP weather models at 10 to 7 days lead time. The former ensemble predicted Sandy's landfall in NY very accurately, while the latter ensemble predicted Sandy going out to sea.\footnote{See, for example, https://www.ecmwf.int/sites/default/files/elibrary/2013/10913-evaluation-forecasts-hurricane-sandy.pdf} Again, for the decision-relevant issue of whether NYC needed to prepare, the models' small structural differences gave totally distinct predictions---the differences in model structure were more important than initial condition uncertainty.

Indeed, in the field of meteorology, it is well accepted that the dominant error growth in weather forecasts is structural. If this were not the case, weather forecasts would not be getting better over time as fast as they are. Some of the improvement is due to better knowledge of initial conditions, but much of it is due to improving model representations of processes. We are still some way off from being able to say that weather forecasts cannot be improved absent better initial conditions.\footnote{See for instance, (Allen et al.,  2002) and (D. Orrell et al., 2001).}

In other words, there are many cases in which model error destroys forecast skill faster than the ordinary or ``classical" chaos inherent in the real-world attractor.\footnote{(Palmer et al, 2014).} These are cases in which model structure uncertainty is more serious than initial condition uncertainty.  Let's call this intuitively characterized state of affairs ``structural chaos" and leave it for now as an intuitive, not-precisely-defined notion. (It must remain somewhat informal since it is defined in pragmatic terms---dependent on the circumstances and aims of the modeler.) This much is clear:  if we confine ourselves to the domain of informal intuitive language, then the question ``is there ever structural chaos in weather and climate modeling?'' can clearly be answered in the affirmative. 

But this leaves open two extremely important sets of questions.  The first is: ``How common is this state of affairs?  Can anything precise be said about a set or sets of models in which this state of affairs is ubiquitous?  Is there, perhaps, a set of models or circumstances in which it in some sense `generic'?"  
%%% measure 1---relative to what space?   EW: isnt that clear in the question?  the "set of models".
The second is: ``Are there precise, formal, mathematical results that could help us settle the first question for us?''  This would obviously require a precise mathematical definition of structural chaos.  Does one exist that is useful here? Failing that, are there lessons to be drawn from the study of toy models about how often to expect these states of affairs in somehow-comparable non-toy systems?  Or, contrarily, \textit{are we forced to examine the empirical evidence on a case-by-case basis to learn which models are sensitively dependent on model structure}? And are we forced to examine the empirical evidence on a case by case basis to learn \textit{which are the circumstances in which model error destroys forecast skill faster than the chaos inherent in the real world attractor}?

Here we turn to some answers to these questions that have been offered by Erica Thompson and further promoted in papers by Roman Frigg, Leonard Smith, and colleagues.\footnote{Hereafter referred to as the ``LSE group''.}\textsuperscript{,}\footnote{(Smith, 2002), (Frigg et al., 2013a), (Frigg et al., 2013b), (Bradley et al., 2014),  (Frigg et al., 2014), (Thompson, 2013).} The LSE group argue that are indeed some very significant mathematical results and go on to answer the above questions in the following way:   With respect to the first set of questions, they claim that these phenomena dominate the space of non-linear models, and that this dominance can be characterized mathematically. They further claim, by way of answer to the second set of questions, that their answers to the first set of questions are secured by mathematical results from the topological study of dynamical systems.  These are results regarding a precise and technical characteristic of dynamical systems known as ``structural stability."  They coin the term the ``hawkmoth effect" to refer both to the absence of structural stability and to what they claim are the epistemological implication of this absence.  Those epistemological implications are roughly as follows:    when structural stability is absent (which, they claim, is nearly everywhere in non-linear modeling), errors in output depend on errors in model structure in a way that is tightly analogous to the phenomenon of sensitive dependence on initial conditions in chaotic systems.  It is not just that the degree of model error we actually have destroys forecast skill faster than classical chaos, according to them, it is that arbitrarily small model structural errors are as dangerous as small errors in initial conditions are in the presence of classical chaos.  It is this alleged tight analogy that leads them to refer to absence of structural stability with a term, ``hawkmoth effect," that is a close cousin of the term ''butterfly effect".  In the presence of uncertain model structure, according to them, therefore, the phenomenon, which is ``generic"\footnote{(Frigg et al, 2014, 45).} in non-linear models, renders mathematical models useless for quantitative prediction. 
The implications of these results, according to them, are  severe. It is a ``poison pill''  that ``pulls the rug from underneath many modeling endeavors''\footnote{(Frigg et al., 2013, p. 479).}. It undermines the quantitative predictive power of almost all non-linear models and makes them incapable of producing ``decision-relevant predictions''\footnote{"If a nonlinear
model has only the slightest SME, then its ability to generate decision-relevant predictions
is compromised."(Frigg et al., 2014, p. 31).} and ``decision-relevant probabilities''\footnote{(Frigg et al., 2013, p. 479).}.  Perhaps most importantly, it shifts the burden of proof\footnote{``[T]he challenge stands: those using nonlinear models for predictive purposes have to argue that the model they use is one that is structurally stable, and this is not an easy task." (Frigg et al., 2014); ``A further possible defence of the default position might just take our results to be inapplicable to actual modelling practices because wether and climate models are very different from our simple logistic map case....We think the burden of proof is on
the proponent of the default position to show what it is that makes such models immune from our criticism.'' (Bradley et al., 2014).} onto anyone who wants to use a non-linear model for anything more than qualitative understanding to demonstrate that this kind of effect is absent, because the effect itself dominates non-linear models.  As they put the point: ``The central claim of this article is that if a nonlinear model has \textit{only the slightest} SME [structural model error], then its ability to generate decision-relevant probabilistic predictions is compromised." (Frigg et al., 2014, p.32, our emphasis)

We think all three of these claims are  overstated. Thus, we challenge three of their central claims: 1) that there are clean and precise mathematical results that are probative regarding the ubiquity of structural chaos (as we have intuitively characterized it) in non-linear modeling (including, of course, climate science);  2) that it is a consequence of these results that a shift in burden of proof is warranted; and 3) that absent such proof, climate models are not ``decision relevant" and present  a number of other epistemically significant problems, even in cases where no empirical evidence for such problems is on offer.  \textit{We conclude that we are forced to examine the empirical evidence on a case-by-case basis to learn which are the circumstances in which model error or model uncertainty destroys forecast skill faster than the chaos inherent in the real world attractor.  We need to look carefully at similar evidence, and at the decision maker's context, to decide what models are decision relevant and can produce decision relevant probabilities even when the likelihood of very small model error is high.}

%%%%%%%%%%%%%%%%%%%%%%%%%%%%%%%%%%%%%%
\section{Structural stability}
%%%%%%%%%%%%%%%%%%%%%%%%%%%%%%%%%%%%%%

\emph{The central claim that we want to rebut is the claim that the ubiquity of structural chaos in non-linear modeling is demonstrable on the basis of either mathematical results or toy model examples.} The mathematical results that the LSE group use to try to secure the claim they cite have to do with a precisely defined topological notion called ``structural stability" and the toy models they appeal to are the logistic model and a very nearby cousin (the ``demon" example.)   

Let us start with their discussion of structural stability.   Before we do, we should be clear that the LSE group use the expression ``the hawkmoth effect'' in two ways.   The first way makes it more or less synonymous with what we have called ``structural chaos" above.  This synonymity is clear from two things:  1) It is clear from what they take to be epistemological consequences of its presence, and 2) it is clear from the fact that they take their ``demon example'' to illustrate its effects. (More on this soon.)

But they also use it in a precise mathematical way, by reference to structural stability.  They use it in these two interchangeable ways because they think that the precise mathematical notion captures well the intuitive one.  We are not convinced of this equivalence.  We do not think absence of structural stability is a form of structural chaos. Since we are not convinced of this equivalence, we use the term ``hawkmoth'' to refer to the absence of structural stability. We use ``structural chaos" to refer to the intuitive notion we characterized above.  And where possible, we try to avoid using the expression ``hawkmoth effect'' since we are not convinced that absence of structural stability leads to something that is well described as an ``effect''.

The LSE group make it clear in a few places that they have in mind a precise mathematical definition of  the hawkmoth effect. The central place  is in (Thompson,  2013). Elsewhere, the reader is simply referred back to Thompson. Indeed, in what might be considered the flagship hawkmoth paper, Frigg et al (2014) describe the hawkmoth effect as follows: ``Thompson (2013) introduced this term in analogy to the butterfly effect. The term \textit{also emphasizes }that structural model error (SME) yields a worse epistemic position than SDIC: hawkmoths are better camouflaged and less photogenic than butterflies." (p.39, emphasis added) They go on to claim that ``what truly limits our predictive ability is not
SDIC but SME." (p.39)

So this passage makes clear what we stated above: that the LSE group have in mind a technical definition of the hawkmoth effect (found in (Thompson, 2013)) as well as an intuitive characterization of it:  it is the structural analog of the butterfly effect, and as we say above, when it is present, model error destroys forecast skill faster than the chaos inherent in the real world attractor.  What is the technical definition?

Looking in (Thompson, 2013) we find the following:

\begin{quotation}
In this chapter I introduce a result from the theory of dynamical systems and demonstrate its relevance for climate science. I name this result, for ease of reference, the Hawkmoth Effect (by analogy with the Butterfly Effect). (p. 211)
\end{quotation}
and
\begin{quotation}
The term “Butterfly Effect” has greatly aided communication and understanding of the consequences of dynamical instability of complex systems. It arises from the title of a talk given by Edward Lorenz in 1972: “Does the flap of a butterfly’s wings in Brazil set off a tornado in Texas?”.

\textit{I propose that the term “Hawkmoth Effect” should be used to refer to structural instability of complex systems.} The primary reason for proposing this term is to continue the lepidoptera theme with a lesser-known but common member of the order. The Hawkmoth is also appropriately camouflaged, and less photogenic. (Thompson, 2013, p. 213, emphasis added.)
\end{quotation}

So, in addition to a definition of `hawkmoth' as absence of structural stability, there are three inter-related claims to address.  The first is that structural stability is very rare in non-linear systems, the second is that its absence more or less guarantees structural chaos, and the last is that it generates epistemological problems that fall under the heading of being a ``poison pill", undermine ``decision relevance'', and shift burdens of proof.  Since ``hawkmothness'' is fundamentally tied to structural stability in this way, the very term ``hawkmoth effect'' only really denotes insofar as the second claim is true.

Given this, the claim that absence of structural stability is akin to model structure chaos is especially important.  We now examine that claim as well as the claim  that structural instability is ubiquitous in non-linear modeling. 

%%%%%%%%%%%%%%%%%%%%%%%%%%%%%%%%%%%%%%
\section{Is absence of structural stability an effect? Is it structural chaos?}
%%%%%%%%%%%%%%%%%%%%%%%%%%%%%%%%%%%%%%

The problem is that we have only given ``structural chaos" an intuitive gloss. It is the idea that model errors destroy forecast skill faster than the chaos inherent in the real world attractor.  In fact, not only is this definition informal rather than mathematical, it is also rather context dependent.  It is a fact about a modeling situation modelers find themselves in, rather than a fact about a model tout court.   So, we need to get quite a bit clearer about what set of formally defined mathematical properties a model could have such that it could be counted on to produce this effect in all, or almost all, modeling contexts.   Here is as good a place as any to start:

The main argumentative thrust of all the LSE group's papers on the hawkmoth effect is that the absence of structural stability is a close cousin of the butterfly effect. Here we probe the cogency of this analogy. 

The more technical term for the ``butterfly effect,'' of course, is ``sensitive dependence on initial conditions'' (SDIC). The definition of SDIC can be formulated in a variety of ways, but two in particular bring out especially salient features. A third definition picks out a related property that is in the family of properties associated with chaotic systems: topological mixing.

\begin{definition}
\label{SDIC delta}
For a state space $X$ with metric $d$, say that the behavior of a dynamical system $(\mathbb{R}, X, \phi)$ with time-evolution function $\phi: \mathbb{R} \times X \rightarrow X$ is sensitive to initial conditions to degree $\Delta$ if for every state $x \in X$ and every arbitrarily small distance $\delta > 0$, there exists a state y within distance $\delta$ of x and a time $t$ such that $d(\phi(t,x),\phi(t,y)) > \Delta$.\footnote{We take this definition almost verbatim from (Werndl, 2009) and (Mayo-Wilson, 2015). One small difference is that we have moved from a definition that applies to maps to one that applies to flows.  (Crudely, a map is a function that we iterate to find a system's trajectory and a flow tells us what happens after a real-numbered value of time.  The difference is discussed more formally below.) Both definition 1 and 2 can be converted from a flow-based definition to a map-based definition with ease. It is worth pointing out that talking about SDIC to degree ``$\Delta$" is very weak since it says nothing at all about how fast you need to get there and since it demands only that some state $y$ near $x$, rather than that almost all states $y$ near $x$, have the property.  We use it for two reasons. One is that we are continuing a conversation that begins with (Frigg et al, 2014) and continues with a response to them from Mayo-Wilson.  The second is that a maximally weak notion of SDIC is maximally favorable to the LSE group since it sets the bar maximally low for them.}
\end{definition}

Informally this says that a system exhibits weak sensitivity to initial conditions if: no matter the true initial state $x$, there is an arbitrarily close state $y$ such that, if $y$ had been the initial state, the future would have been radically different (to the degree $\Delta$).  We could also strengthen the definition of sensitivity to initial conditions to require that almost all such states have this property. 

\begin{definition}
\label{SDIC exponential}
For a state space $X$ with metric $d$, say that the behavior of a dynamical system $(\mathbb{R}, X, \phi)$ is exponentially sensitive to initial conditions if there exists $\lambda > 0$ such that for any state $x \in X$, any $\delta > 0$, and all times $t$, almost all elements $y \in X$ satisfying $0 < d(x,y) < \delta$ are such that $d(\phi(t,x),\phi(t,y))>e^{\lambda t}d(x, y)$.\footnote{ Note that this definition talks of ``almost all" states, without there being specific mention of a measure.  This is fairly standard; the reader is free to interpret them as either conditional on a specified metric or, as we more naturally intend it, as presupposing the Lebesgue measure, a standard practice in discussions of the state space of classical systems.  Of course this is a significant problem, as we will see, for the LSE group, since there is no similarly natural measure on the space of evolution functions.  Finally, just as the reader can easily convert either definition back and forth between a map and a flow, she can also convert them back and forth from being what is sometimes called a ``strong" version (where the claim is about almost all nearby states) and a ``weak" version (where the claim is about at least one nearby state). We have chosen to follow Mayo-Wilson and Werndl in giving Definition 1 a ``weak" form but we have given definition 2 a ``strong" form.  Only strong versions of such definitions, obviously, require a measure, but only strong versions are usually taken to have strong epistemological consequences, since they are \textit{likely} to produce error.}
\end{definition}

Intuitively, not only does exponential SDIC allow you get to arbitrarily far away (subject to the boundedness of your dynamics) from where you would have gone by changing your initial state just a very small amount, but this definition requires that you be able to get there very fast. More precisely, it says that there will be exponential growth in error---that every  $1 /\lambda$ units of time (called the ``Lyapunov time'') the distance between the trajectories picked out by the two close-by initial states will increase by a factor of $e$.  We assume that exponential SDIC is  ``strong'': that it also requires that almost all the points near $x$ take you there, not just one.  Strong, exponential SDIC is what people usually have in mind when they talk about the butterfly effect.

\begin{definition}
\label{topological mixing}
A time-evolution function $\phi$ is called topologically mixing if for any pair of non-empty open sets $U$ and $V$, there exists a time $T > 0$ such that for all $t > T$, $\phi(\{t\} \times U) \cap V \neq \emptyset$.
\end{definition}

Informally, topological mixing (a crucial ingredient of chaos) occurs if, no matter how arbitrarily close one starts, one will eventually be driven anywhere in the state space that one likes. 

It stands to reason, then, that if the Miller analogy holds, then absence of structural stability should involve a cluster of properties of dynamical systems that parallels the above three, but where the notion of two states being  close is replaced with the notion of two \textit{equations of evolution} being close.  We will call making this kind of replacement a ``lepidopteric permutation" of a definition associated with the butterfly effect because the hawkmoth and the butterfly are both Lepidoptera. 

Definition \ref{topological mixing}  is the easiest of the three on which to perform the lepidopteric permutation, because it is topological, and topology is the natural domain in which to study families of dynamical systems (whose members do not have a natural metric between them). In fact, Conor Mayo-Wilson\footnote{(Mayo-Wilson, 2015) Our definition 3* is similar to and inspired by his attempt to capture one aspect of what the LSE group might mean by the hawkmoth effect, but we also emphasize that topological mixing is only one aspect of chaos.  It happens not to be the feature of chaos, moreover, that is usually associated with the butterfly effect. And finally, in so far as one is looking for the lepidopteric analog of topological mixing, which is a purely topological notion, definition 3** makes the most sense, since it is also topological.} has proposed a reasonable candidate for a hawkmothified version of definition 3. In parallel with the notion of topological mixing, he calls it structural mixing. We do something similar by taking definition 3 and replacing the neighborhood of initial states with a neighborhood of evolution functions:

\begin{cdefinition}{3*}
\label{structural mixing}
Let $\Phi$ be a space of time-evolution functions with metric $\delta$, $\phi\in \Phi$, $U$ be a non-empty open subset of $X$, and $\epsilon > 0$. Furthermore, for any time $t$, set $B_{\epsilon}(\phi)(t,x) = \{\phi'(t,x) \mid \delta(\phi, \phi') < \epsilon\}$. We say that $\Phi$ is structurally mixing at $\phi$ if for any state $x \in X$, there is a time $T$ such that for all $t > T$, $B_{\epsilon}(\phi)(t,x) \cap U \neq \emptyset$.
\end{cdefinition}

This definition uses a metric to pick out the preferred topology on the space of evolution functions, but it makes more sense to relax it to an arbitrary topology.

\begin{cdefinition}{3**}
\label{structural mixing revamped}
Let $\Phi$ be a space of time-evolution functions, $\phi\in \Phi$, $U$ be a non-empty open subset of $X$, and $V$ be a non-empty open subset of $\Phi$ (in the appropriate topology) containing $\phi$. Furthermore, for any time $t$, set $V(t,x) = \{\phi'(t,x) \mid \phi' \in V \}$. We say that $\Phi$ is structurally mixing at $\phi$ if for any state $x \in X$, there is a time $T$ such that for all $t > T$, $V(t,x) \cap U \neq \emptyset$.
\end{cdefinition}

Definition \ref{SDIC delta} can also be finessed into a structural equivalent. Now we absolutely need a metric of distance between two evolution-specifying functions. We are of course free to pick one, but it is worth keeping in mind that in many cases there will be no natural choice.   We would get a notion of ``sensitive dependence on model structure to degree $\Delta$'' and define it roughly as follows (we use the ``strong" version of Definition 1 since that is the epistemologically interesting version of SDIC):

\begin{cdefinition}{1*}
\label{sdms}
Let $\Phi$ be a space of time-evolution functions with metric $\delta$, $\phi\in \Phi$, and $\epsilon >0$. We say that $\Phi$ is sensitively dependent on model structure to degree $\Delta$ at $\phi$ if for any state $x \in X$, there is a time $t$ such that for almost all  $\phi' \in \Phi$ satisfying $\delta(\phi,\phi') < \epsilon$, we have $d(\phi'(t,x),\phi(t,x)) > \Delta$.
\end{cdefinition}

The reader can easily construct a weak version of this for herself.

Definition \ref{SDIC exponential} is a definition on which it is very hard to see  how to perform a lepidopteric transformation, because Definition \ref{SDIC exponential} gives a requirement in terms of an exponential growth in a single quantity: the  distance of separation between nearby states, as the system evolves in time. We are already brushing under the rug the fact that \ref{sdms} and \ref{structural mixing} are assuming the existence of a metric of distance between evolution functions. But in a hawkmoth version of Definition \ref{SDIC exponential}, unlike in the butterfly version, there is no single quantity that can grow in time. A hawkmoth version would have to relate model distance with state space distance.  It would have to coordinate a metric of model structure distance with a metric of state space distance. This will turn out, we will see, to actually be even more challenging than it seems.

That the two above definitions should at least be considered to be in the right neighborhood of what a hawkmoth effect should involve is well motivated by the butterfly/hawkmoth analogy. But we needn't put too much weight on this analogy.  

\begin{figure}[ht]
\begin{center}
\includegraphics{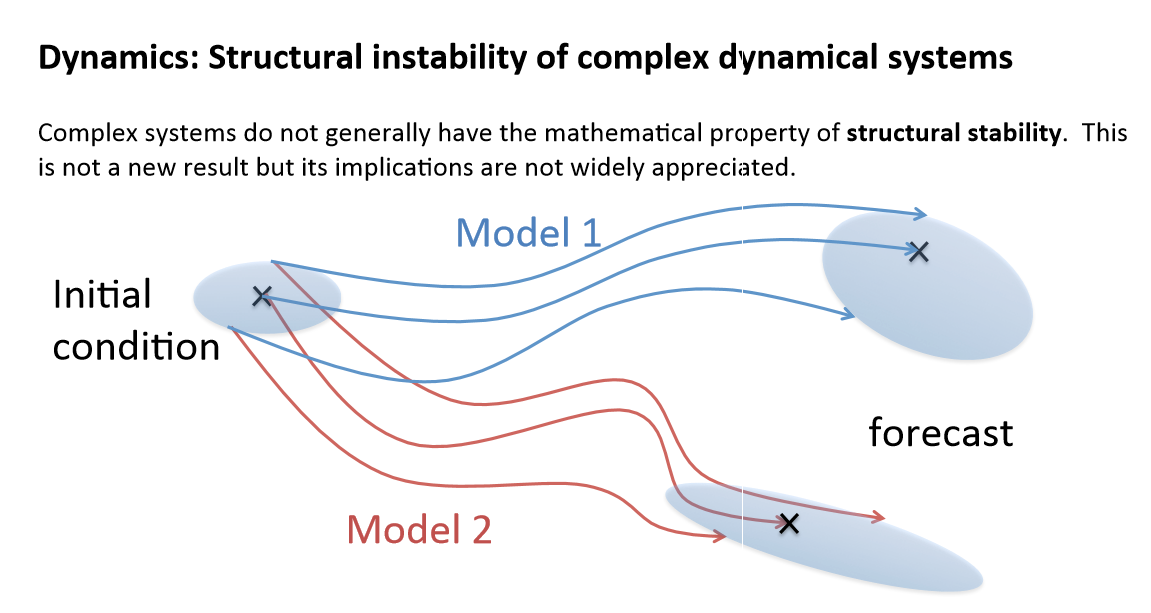}
\caption[]{A visual representation of the hawkmoth effect, or lack of structural stability, from Thompson and Smith.\footnotemark}
\label{fig:hawkmoth}
\end{center}
\end{figure}

We can also look at some of the claims that they make about the epistemological consequences of a hawkmoth effect or absence of structural stability. They claim that 1) the hawkmoth effect ruins the closeness-to-goodness rule, which suggests that a close-by model is a reasonably good model for producing forecasts,\footnote{``This distribution, so the argument goes, has been carefully chosen to drive our point home, but most other distributions would not be misleading in such a way, and our result only shows that unexpected results can occur every now and then but does not amount to a wholesale rejection of the closeness-to-goodness link. There is of course no denying that the above calculations rely on a particular initial distribution, but \textit{that realization does not rehabilitate the closeness-to-goodness link}." (Frigg et al., 2014, p.40, our emphasis).} 2) the hawkmoth effect causes a fast growth in entropy,\footnote{``we expect [nearby models] to have growing relative entropy" (Frigg et al., 2014, p.47).} and 3) the hawkmoth effect is an underappreciated cousin of the butterfly effect with similar epistemological consequences.\footnote{``In this article we draw attention to an additional problem than has been underappreciated, namely, structural model error (SME)...SME in fact puts us in a worse epistemic situation than SDIC.'' (Frigg et al., 2014).} And they show pictures like Figure \ref{fig:hawkmoth}.  If the hawkmoth effect does not have this cluster of properties, it is hard to see why anyone should think  that it poses the kinds of epistemological challenges that the LSE group claims it does.\footnote{Or that the ``demon'' case  we discuss below is a reasonable illustration of it.}

\addtocounter{footnote}{-1}

\footnotetext{Source: http://www.lse.ac.uk/CATS/Talks and Presentations/Posters/Thompson-TheHawkmothEffect-LSEResearchFestival2014.pdf.}

It should be reasonably clear that Figure \ref{fig:hawkmoth} is a good depiction of what we have called sensitive dependence on model structure to degree $\Delta$, as defined in \ref{sdms}, and vice versa---if $\Delta$ represents roughly the distance between the two blue sets on the right.  

All of this is further reason to think that if a precise mathematical definition of the hawkmoth effect exists, and if that definition is going to capture both the analogy between the hawkmoth and the butterfly, as well as secure that claim that it results in cases in which model error destroys forecast skill faster than the chaos inherent in the real world attractor, then that mathematical definition is going to have to implicate something like a cluster of properties that are the analogs of the 3 properties, above, that we associated with the butterfly effect and with chaos.  We note, however, that we have already encountered a conceptual problem because a relevant analog of property 2 is lacking.

Let us now look and see whether or not absence of structural stability implicates anything like these 3 properties. Interestingly, the term ``structural \textbf{in}stability'' doesn't seem to appear very much in the mathematical literature.  Nowhere is it described as producing an ``effect.'' In fact, the discussion that we find in both (Thompson, 2013, ``5.2.2 Identifying structurally stable systems'', p. 214-215), and even more so (Frigg et al., 2014, ``4. From Example to Generalization'', p. 45-47) are clearly influenced by (Pugh and Peixoto, 2008).  But that source does not use the term ``instability'' nor does it describe an ``effect''.

Why does this matter? It matters because structural stability does not have a  complement with substantial features of its own, and the term ``structural instability'' suggests an overly close analogy to chaotic instability that no one in the mathematical literature ever had in mind. We can see why if we look closely at the notion of  structural stability.

There are a few different definitions of structural stability that we can find in the literature.  One good survey can be found in (Pugh and Peixoto, 2008). The first thing we should notice is that they are definitions of ways of guaranteeing two trajectories to stay arbitrarily close. And failure to stay arbitrarily close is not the same thing as being guaranteed to go an arbitrary (bounded) distance away. But in analogizing absence of structural stability to SDIC, the LSE group are engaging in exactly this conflation.

Take an early definition of structural stability in two dimensions due to Andronov and Pontrjagin as it is explained in Pugh and Peixoto. They consider dynamical systems of the form
\begin{equation}
\frac{dx}{dt}=P(x,y), \quad \frac{dy}{dt}=Q(x,y) 
\end{equation}
defined on the disc $D^2$ in the $xy$-plane, with the vector field $(P,Q)$ entering transversally across the boundary $\partial D^2$. 

\begin{definition}
\label{And and Pon}
Let $p$ and $q$ be vector fields on $D^{2}$. We say that such a system is structurally stable (or ``rough'', as Andronov and Pontrjagin put it) if, given $\epsilon > 0$, there exists $\delta > 0$ such that whenever $p(x,y)$ and $q(x,y)$, together with their first derivatives, are less than $\delta$ in absolute value, then the perturbed system 
\begin{equation}
\frac{dx}{dt}=P(x,y) + p(x,y), \quad 
\frac{dy}{dt}=Q(x,y) + q(x,y) 
\end{equation}
is such that there exists an $\epsilon$-homeomorphism $h:D^2 \rightarrow D^2$ ($h$ moves each point in $D^2$ by less than $\epsilon$) which transforms the trajectories of the original system to the trajectories of the perturbed system.
\end{definition}

Intuitively, this definition says of a particular evolution function that no matter how $\epsilon$-close to the trajectory of that evolution function I want to stay \textit{for its entire history}, I can be guaranteed to find a $\delta$ such that all evolution functions within $\delta$ of my original one stay $\epsilon$-close to the original trajectory---where $\delta$ is a measure of how small both the perturbing function and their first derivatives are. This is an incredibly stringent requirement.

When an evolution function fails to obtain such a feature, therefore, it is as misleading to call it ``structurally unstable'' as it is to talk about a system being insensitively dependent on initial conditions, or to describe a function that fails to have a particular limit as ``unlimited.'' It is an ambiguity that conflates the following two sorts of claims.

\begin{enumerate}
\item One cannot be guaranteed to stay arbitrarily close by choosing an evolution function that is within some small neighborhood.
\item Small changes in the evolution function are sure to take one arbitrarily far away (subject to the boundedness of the system).
\end{enumerate}

\begin{figure}[ht]
\begin{center}
\includegraphics[scale=.12]{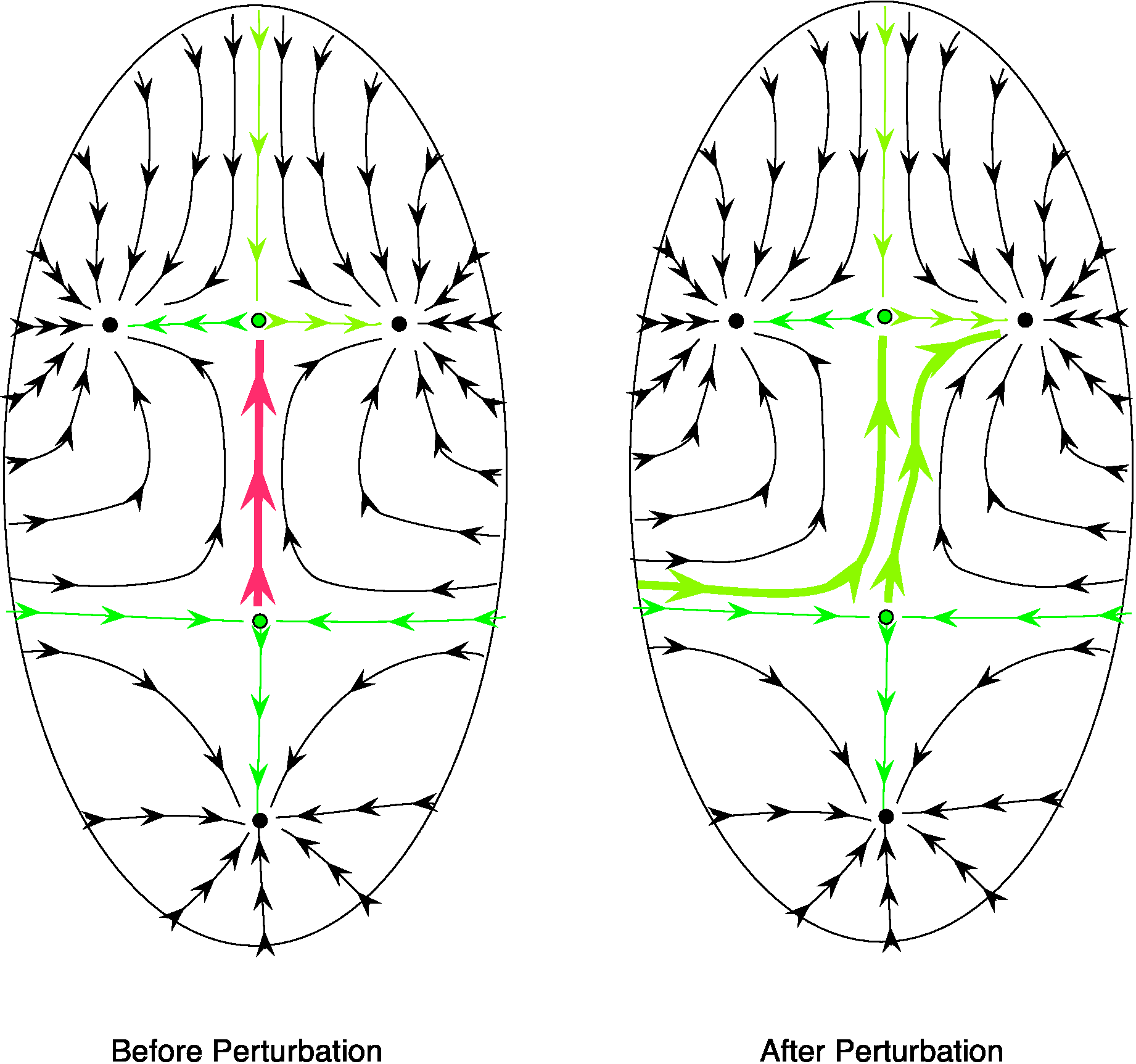}
\caption{An example of a map that is not structurally stable.  When a small perturbation turns the map on the left into the map on the right, we get two maps that cannot be smoothly deformed into each other.  The key feature is this qualitative dissimilarity between the two maps, and not any metrical difference. (Source: Pugh and Peixoto, 2008)}
\label{fig:structural stability picture}
\end{center}
\end{figure}

Absence of structural stability in the sense of definition \ref{And and Pon} gives you the first thing, but nothing anywhere near approximating the second thing. But to claim that absence of structural stability is an analog of SDIC is to suggest that absence of structural stability implies the second thing. It is the second thing, moreover, that we concluded that a hawkmoth effect should ensure when we formulated definitions \ref{sdms} and \ref{structural mixing} above using the lepidopteric transformation of definitions \ref{SDIC delta} and \ref{topological mixing}.
\label{discussion of measure}
That, moreover, is just the first difference.  Notice that  \ref{sdms}, in order to be the analog of strong SDIC, has to say ``for almost any  $\phi \in \Phi$''.  But absence of structural stability in the Andronov and Pontrjagin sense requires nothing of the sort.  It only requires that, for each $\delta$, one trajectory in the entire set $\delta$-close trajectories be deformed by more than $\epsilon$ (see Figure~\ref{fig:structural stability picture}).

To put the point simply, absence of structural stability in the Andronov and Pontrjagin sense is much weaker than either definitions \ref{sdms} or \ref{structural mixing}.  Rather than requiring that \textit{most} trajectories (indeed as we have seen there is no coherent notion of ``most" here) go \textit{far} away, it only requires that \textit{one} trajectory go \textit{more than a very small epsilon} away.  And hence it is much weaker than the LSE group or the hawkmoth analogy suggest. This results from taking the complement of a definition. Definition \ref{And and Pon} is ``strong.'' It requires that \textit{not a single trajectory} diverge by more than $\epsilon$.  But of course, the complement of a strong definition is a weak one.  Here, the complement of structural stability only requires that, for each closeness threshold $\delta$, one trajectory go astray.

Compare absence of structural stability to SDIC.  In SDIC, given an arbitrarily large (finite) distance, $\delta$, and another arbitrarily small (finite) distance, $\epsilon$, there exists a time $T$ such that the dynamics will take balls of size $\epsilon$ to balls of size $\delta$ after time $T$.  But one cannot show absence of structural instability by picking a large delta to start.  Without picking a small enough ball to start with, lack of structural stability would not guarantee to get one out of the ball, no matter how long the wait, and no matter how large the initial ball is.  It only guarantees that if one picks a small enough $\epsilon$, then one can eventually get out of the epsilon-sized ball.  This is a mathematical fact that is independent of your domain of application.    Lack of structural stability simply has no consequences for prediction on any time scale, or for any metric of significance of predictive error.

Let us now turn to the claims about the likelihood of an arbitrary model being structurally unstable.  In reading their papers and in conversation, one sometimes gets the impression that the LSE group think a mathematical result in topology due to Smale can stand in for a claim about the likelihood of a system being structurally stable.  But this is at best a very weak argument and at worst a confusion. They summarize Smale's result as follows: ``Smale, 1966, showed that structural stability is not generic in the class of diffeomorphisms on a manifold: the set of structurally stable systems is open but not dense.''   But Smale himself certainly never uses the term ``generic," nor any term that we would regard as being a close cousin.  And density is a topological notion with no obvious measure-theoretic implications.  Unfortunately, his is the only argument we can find in the results they review about structural stability that they use in support of the claim that ``there are in fact mathematical considerations suggesting that the effects we describe are generic." (Frigg et al., 2014, p. 57)  

We think this might be a confusion because it is not clear to us how any of the mathematical consideration surveyed suggest these things.  For one, density is not a measure-theoretic notion, it is a topological one.  A set can be dense and have measure zero (think of the rationals in the real number line---the rationals are of course not generic in the real line). There are even nowhere-dense sets that have arbitrarily high measure in the reals. There is a general point here: all the relevant notions associated with structural stability are topological, and they provide no information about likelihoods, or `generic'-ness.  Moreover, the effects that the LSE group emphasize are structural instability, not stability. Even if we assume some relevant non-measure-theoretic notion of genericness, the first result they describe, due to Smale, would show that structural stability \textit{is} generic in certain settings. Another result of Smale's shows that structural stability is not generic in another setting, but that hardly entails that instability is generic. A third result they mention, from Judd and Smith, does not mention structural stability at all. The LSE group claims they are giving \textit{plausibility} arguments, but it is hard for us to see how what they say amounts to a plausibility argument.

We still have not discussed, moreover, definition \ref{SDIC exponential} of SDIC and the idea of exponential error growth, which is so fundamental to the epistemological impact of chaos. In fact, even more than \ref{sdms} and \ref{structural mixing} would be needed to underwrite the existence of a hawkmoth effect.  We would need to formulate a definition that captured the idea that the small error in a model could grow very fast---indeed we would need something akin to definition \ref{SDIC exponential} that would allow us to calculate how soon a forecast would become useless given a certain amount of model structure uncertainty.  No other way will be able to make precise the idea that \textit{model error destroys forecast skill faster than the chaos inherent in the real world attractor}. But this looks unlikely for the case of failure of stability in the sense of Andonov and Pontrjagin. The reason is that in definition \ref{And and Pon}, the metric of model distance does not live in the same space as where the metric of state space distance lives. It would be strange and confusing to relate these two metrics in a single equation.

Things get even worse if we move from what Pugh and Peixoto call the ``pre-history'' of structural stability to its modern formulation.  In the modern formulation, no metric is specified---neither between two different diffeomorphisms, nor between two trajectories.  

The modern formulation of structural stability goes as follows:
\begin{definition}
\label{structural stability of a map}
If $D$ is the set of self-diffeomorphisms of a compact smooth manifold $M$, and D is equipped with the $C^1$ topology, then $f \in D$ is structurally stable if and only if for each $g$ in some neighborhood of $f$ in $D$ there is a homeomorphism $h: M \rightarrow M$ such that $f$ is topologically conjugate to each nearby $g$. In other words, that 
\begin{equation}
\begin{CD}
M @> f >> M \\
@VhVV @VVhV \\
M @> g >> M
\end{CD}
\end{equation}
commutes, that is, $h(f(x)) = g(h(x))$ for all $x \in M$.
\end{definition}
Definition \ref{structural stability of a map} makes it clear that the most general formulation of it is not metrical at all. It is topological. It says nothing at all about diffeomorphisms that are ``a small distance away''. It talks about diffeomorphisms that are in topological neighborhoods of each other. And it does not at all mention trajectories taking you some distance away. It talks about there being a homeomorphism (a topology-preserving transformation that cares nothing about distances) between the two trajectories.  Using the analogy of a rubber sheet that is often used to explain topological notions, it roughly says, intuitively, that if you replace $f$ by any of the diffeomorphisms $g$ in some neighborhood of $f$ then the new entire statespace diagram of $g$ will be one that could have been made just by stretching or unstretching (by deforming it in the way one can deform a rubber sheet without tearing it) the statespace diagram of $f$. 

The main point here is this:  the epistemological punch of the butterfly effect results comes from being able to show that certain quantifiable deficiencies in our predictive capacities arise from the butterfly effect.  But absence of structural stability does not provide these deficiencies.   The results are just too far from anything remotely like a metric of error.

So let's review what it amounts to for a system defined by a diffeomorphic map $f$, applied to a manifold, to fail to be structurally stable.  It means that if you look at the space of diffeomorphisms around $f$, you will be unable to find a neighborhood around $f$ that is guaranteed not to contain a single other diffeomorphism $g$ that is qualitatively different than $f$ in essentially the following sense:  that you cannot smoothly deform $f$ into $g$.  The Father Guido Sarducci version of what we have learned so far is this: \textit{You don't get structural `instability' just by replacing ``small changes in initial conditions" in SDIC with ``small changes in model structure" because, both with regard to how much error you can get, and with respect to how many nearby trajectories will do it, SDIC says things are maximally bad, while structural instability merely says they will not be maximally good.  And SDIC includes metrical claims which are lacking in the case of structural stability.}

\section{The epistemological consequences of absence of structural stability, in a nutshell}

Let us put this another way. Absence of structural stability fails to have the right sort of properties to raise the kinds of skeptical worries that the LSE group claims it does for three different reasons.   First and foremost, it is a property that relates to topological features of a family of models, not metrical ones.  But very tiny changes (from a metrical point of view) can create fundamental topological differences.  Unlike the butterfly effect, structural stability says nothing at all about errors \textit{becoming large}.   Second, structural stability fails to obtain if there is a single model in your neighborhood whose future evolution destroys the topology.  This is a strong constraint on the epistemological significance of absence of structural stability:  you could know for a fact that your model is structurally unstable, know for a fact that you had some small amount of stuctural model error, and still have it be the case that your model would not introduce more than an arbitrarily small amount of error for an arbitrarily long time.  And it could still be overwhelmingly likely (no matter what measure you preferred on the model space!) that it would introduce virtually no error at all. This is all of course because structural stability was a notion developed by people interested in achieving the \textit{mathematical certainty of proof while using perturbations}---they were not interested in finding predictive accuracy.  After all, they were studying the solar system.  No one was worried that, until they found stability results for the solar system, its dynamical study would be embroiled in confusion or maladapted to quantitative prediction.  Absence of structural stability has significantly weaker implications for reliable predictive accuracy than the butterfly effect.\footnote{It might have some.  A reviewer notes, for example, that different models of the Earth's orbital parameters make a big difference in the predictability of Milankovitch cycles back in time - which has relevance for paleo-climate dating and interpretation.}  Many structurally unstable systems are predictively extremely well-behaved.  The structural instability of the solar system does not make our models of it maladapted for prediction.  Nor, by the way, do any of the results having to do with structural stability make any mention at all of non-linearity.  Non-linearity is a red herring vis-a-vis structural stability.

Absence of structural stability does not deserve the name ``hawkmoth effect".  It is not a form of model structure chaos analogous to the butterfly effect. It does not imply that model error destroys forecast skill faster than the chaos inherent in the real world attractor. And it is not ubiquitous in non-linear models.

We should acknowledge that none of this is intrinsically newsworthy.  But we believe it is worth clearing all of this up given some misleading claims that have been made in the philosophical literature:  that the absence of structural stability is in any way analogous to a butterfly effect;  that it (in anything like the normal cases) does something akin to what we see in figure 1; that it undermines the predictive capacity of nonlinear science; and  that it undermines the capacity of the same to produce decision relevant probabilities.  And that applies, inter alia, to the way in which that is done \textit{both} in weather prediction and in climate projection.  As we will see, it is also misleading to suggest that the LSE group's famous demon's apprentice example in any way illustrates the typical effects of the absence of structural stability.

%%%%%%%%%%%%%%%%%%%%%%%%%%%%%%%%%%%%%%
\section{Putting the demon's apprentice in context}
%%%%%%%%%%%%%%%%%%%%%%%%%%%%%%%%%%%%%%

In addition to canvassing some results regarding structural instability, the LSE group use, in several papers, a toy example of using a non-linear model to produce a probability distribution that they call the "demon's apprentice" example.  It involves comparing the output of the logistic equation to what they claim is another, nearby, map.   

So what about this `demon's apprentice' example?  Does it not provide its own cautionary tale, irrespective of the epistemological import of structural stability considerations?   Does it not do this given that figure 1 \textit{does} seem to capture well what happens in the demon case? Does it not do this given that it seems to show that very nearby models can take a probability distribution over some relatively small set of initial conditions and very quickly drive that set into very different regions of state space?  We turn to these questions in the next section.

In  (Frigg et al., 2014), the LSE group postulate the existence of a demon that is omniscient regarding the exact initial conditions of a given system, the true dynamical model of the system, and the computational output of such a model at any future time, for any initial conditions, to arbitrary precision.  Such a demon also has two apprentices: a senior apprentice, who has omniscience of computation and dynamics yet lacks that of initial conditions, and a freshman apprentice, who has computational omniscience but not that of model structure nor initial conditions.

The problems of the senior apprentice can be overcome by Probabilistic Initial Condition Ensemble Forecasting (PICEF).  In PICEF, instead of using a single point in the state space as the initial conditions for the dynamical system, a probability distribution is substituted over the entire state space. In this way, the practical limitations of initial condition uncertainty can be mitigated: A point prediction for the state of the system in the future is replaced by a probability distribution over possible future states which may still inform practical considerations.

According to Frigg et al., however, there is no such solution for the freshman's ignorance.  Aside from the initial condition uncertainties which reduce the output precision of a system's trajectories, the freshman apprentice is beset by unreliability in the very dynamics by which initial states are evolved.  This unreliability, they claim, is not easily resolved nor easily dismissed.  And its consequences, they hold, are severe.

To illustrate this severity,  Frigg et al. consider the logistic map, defined below, and a ``similar" equation that represents the true model of some physical system.  They show that, given enough time, the two equations evolve the same distribution of initial conditions to very different regions of state space. 
We have already seen that the absence of structural stability is not, in general, as severe as the results of the demon's apprentice example suggest.  Absence of structural stability does not generally lead to wide divergence, nor does its absence imply anything  about the majority of nearby models (see section 4). It only takes one stray model in the neighborhood to violate the definition.   But we can still ask whether the demon example is at least \textit{a} possible illustration of the absence of structural stability.  And we can still ask if it provides a worthwhile cautionary tale of its own. The answers to both of these questions is ``no."

Why is the demon example not a possible illustration of the absence of structural stability?  To see why, we need to review some conceptual distinctions relevant to dynamical systems.   We can start with the logistic equation, which is part of the demon example:
\begin{equation}
\label{logistic1}
x_{t+1} = 4x_{t}(1 - x_{t}).
\end{equation}

Notice that this is a dynamical system specified with an equation of evolution that lives in discrete time. This needs to be contrasted with dynamical systems specified with time-dependent differential equations, like the Lorenz model
\begin{equation}
\label{lorenz}
\frac{dx}{dt} = \sigma (y - x), \quad
\frac{dx}{dt} = x (\rho - z) - y, \quad
\frac{dz}{dt}= x y - \beta z.
\end{equation}

In the formalisms used to discuss structural stability, dynamical systems like the logistic equation are specified by maps: functions from a manifold onto itself; and those like the Lorenz model by flows: a map from a manifold (state space) crossed with the real number line (a time variable) onto the manifold: $\phi : M \times \mathbb{R} \rightarrow M$.

Definition \ref{structural stability of a map} gave us the definition of structural stability for a map, and the definition of structural stability for a flow is:

\begin{definition}
\label{structural stability for a flow}
If $X$ is the set of smooth vector fields on a manifold M equipped with the $C^1$ topology, then the flow generated by $x \in X$ is structurally stable if and only if for each $y$ in the neighborhood of $x$ in $X$ there is a homeomorphism $h:M \rightarrow M$ that sends the orbits of $x$ to the orbits of $y$ while preserving the orientation of the orbits.
\end{definition}

But notice that Definition \ref{structural stability of a map} does not apply to any old map. The definition only applies if the map is a diffeomorphism. To be a diffeomorphism, a map has to have some added conditions:

\begin{enumerate}
\item The map has to be differentiable.
\item The map has to be a bijection (its inverse must also be a function).
\item The inverse of the map has to be differentiable.
\end{enumerate}

The obvious problem is that the logistic map is not a bijection! Every number other than 1 has two preimages. For example, both .3 and .7 map to .84. So .84 has no unique preimage and there is no function that is the inverse of equation \ref{logistic1}. But this means that the logistic map isn't even the right category of object to be structurally stable or not.\footnote{There are other nearby puzzles:  the best model climate science could write down---that is the real true, partial-differential-equation-specified, undiscretized model---would have the form $\phi : M \times \mathbb{R} \rightarrow M$.  It might or might not meet the additional criteria for being a flow, but it is certainly not of the form $\phi : M  \rightarrow M$, which is the general form of a map.  Once we start to think about a discretized model, however, the model does take the form $\phi : M \rightarrow M$.  Even if we had the perfect climate model and it \textit{were} a flow, a discretization of it (in time) would necessarily have the form of a map.  And no map is in the right universality class---for purposes of structural stability---of a flow.  In the sense relevant to structural stability, its simply a category error to ask if a discretization of a dynamical system of the form $\phi : M \times \mathbb{R} \rightarrow M$ is ``nearby to" the undiscretized system.  Climate models run on computers are all imperfect, but they don't live in the same universe of functions as the ``true'' model of the climate does.  Its of course true that there are no guarantees that any currently realizable model ensemble will produce realistic pdfs for a very different globally warmed planet. After all, what if there are missing processes that only kick in at higher temperatures?  There are lots of reasons why discretization can lead to errors. But the only issue \textit{here} is whether mathematical results from the field of structural stability are applicable.  And they are only applicable when certain formal preconditions obtain.  Those do not appear to obtain here.}
Of course we are free to make up our own definition of structural stability that applies to all maps. But if we do, if we expand our model class beyond the space of diffeomorphisms to the space of all maps, then the very notion of structural stability becomes empty.  You simply won't find many structurally stable maps on this definition.  Consider the simplest map there is:

\begin{equation}
\label{constant}
f(x_{i}) = C  \quad \textrm{for all } i \textrm{ and some constant } C.
\end{equation}
This map is simple but, of course, not a bijection.  Yet it also would not come out as ``structurally stable'' on our new definition. According to the definition we require all $g$ in some neighborhood of $f$ to satisfy $h(f(x)) = g(h(x))$. But if $f$ is constant, $h(f(x))$ is constant, so $g(h(x))$ has to be  constant for the definition to hold. Yet that condition is easy to violate with many of the $g$'s in any neighborhood of $f$.  This implies that the logistic map is no more ``structurally unstable" than the function given by equation \ref{constant} is. Which means whatever the demon example illustrates, it actually has nothing at all to do with non-linearity. (Technically, of course, equation \ref{constant} is not linear.  But it is also not exactly what people have in mind when they think of non-linearity!)  And of course, once you open up the model class to non-diffeomorphisms, $f(x) = 3x$ (an obviously linear map) will also be structurally unstable.\footnote{An anonymous referee suggest that there are competing definitions of structural stability for maps that do not include the diffeomorphism requirement.  Robert Devaney, for example, defines a notion of $C^2$ structural stability that does not require a map to be a bijection.  But Devaney also shows that the logistic equation is $C^2$ structurally stable for many parameter values.  So adopting such a notion of structural stability does not clearly make the problem of the logistic equation being a poor illustration of absence of structural stability go away entirely.  In any case, it is not clear what the logistic equation could have to do with more traditional mathematical definitions of structural stability of the kind the LSE group discuss and regarding which the theorems they reference are about.}

So when Frigg et al (2014) write, ``The relation between structural stability and the Demon scenario is obvious: if the original system is the true dynamics, then the true dynamics has to be structurally stable for the Freshman’s close-by model to yield close-by results,"(p.47) this overstates the case.  In fact the logistic equation is not even a candidate for structural stability.  And as we have seen a model needn't be  structurally stable for a nearby model to produce nearby results.    

Still, even if the logistic map is not a candidate for structural stability, one might claim that the demon example still shows that two very nearby models can lead to radically different PICEF predictions. We saw in section 3 that structural stability and its absence did not underwrite what is depicted in figure \ref{fig:hawkmoth}. But surely the demon example does, right? This, after all, we can see with our own eyes in the Frigg et al (2014) paper. Not quite; here we explain why the notion of ``nearby" models in the demon example is suspect. Let us look carefully at the two models in the example: the freshman apprentice's model and the demon's model.

\begin{equation}
\label{logistic2}
x_{t+1} = 4x_{t}(1 - x_{t})
\end{equation}
\begin{equation}
\label{modified_logistic}
\tilde{x}_{t+1} = (1 - \epsilon)4\tilde{x}_{t}(1 - \tilde{x}_{t}) + \frac{16\epsilon}{5}\left[ \tilde{x}_{t}(1 - 2\tilde{x}_{t}^{2} + \tilde{x}_{t}^{3}) \right]
\end{equation}

Equation \ref{modified_logistic} is the demon's and senior apprentice's model, the ``true'' model in this scenario, and equation \ref{logistic2} is the freshman apprentice's model, the ``approximate'' model.

On first glance, these equations do not look very similar.  But the LSE group argue that they are in fact similar.  They argue this by arguing that the appropriate metric of similarity should be based on an output comparison of the two models over one timestep: Call the maximum difference that the two models can produce for any arbitrary input $\epsilon$, the maximum one-step error.  If $\epsilon$ is sufficiently small, then the two models can be said, according to the LSE group, to be very similar.  Here, we argue that this is much too weak of a condition for considering two models similar. {We also note, as we already did above, that it is not for nothing that the modern literature on structural stability is topological and not metrical.  There simply is not anything sufficiently general and sufficiently natural to say about how to measure the distance between two models, two diffeomorphisms, or two flows.}  The model of equation \ref{logistic2} and the model of equation \ref{modified_logistic} are not appropriately similar for drawing the conclusions that the LSE group draw. We begin by proving a theorem (proofs of theorems can be found in the Appendix):

%%%%%%%%%%%%%%%%%%%%%%%%%%%%%%%%%%%%%%
\section{Small arbitrary model perturbations}
%%%%%%%%%%%%%%%%%%%%%%%%%%%%%%%%%%%%%%

Suppose we are given a difference equation of the form 

\begin{equation}
\label{3}
x_{n+1} = f(x_{n}),
\end{equation}
where $x_{i} \in \mathbb{R}$ and $f: A \rightarrow B$ is an arbitrary function from the bounded interval $A \subset \mathbb{R}$ into the bounded interval $B \subset \mathbb{R}$. Note that the logistic map takes this form with $f(x_{n}) = 4x_{n}(1 - x_{n})$. Then we have the following result:

\begin{ctheorem}{1}
\label{small_model_perturbations}
Given any function $g: A \rightarrow B$ and $\epsilon > 0$, there exists $\delta > 0$ and $\eta > 0$ such that the maximum one-step error of
\begin{equation}
\label{10}
x'_{n+1} = \eta f(x'_{n}) + \delta g(x'_{n}),
\end{equation}
from (\ref{3}) is at most $\epsilon$ and $x'_{n+1} \in B$.
\end{ctheorem}

Observe that (\ref{modified_logistic}) takes the form of (\ref{10}) with $f(x'_{n}) = 4x'_{n}(1 - x'_{n})$, $g(x'_{n}) = (16/5)[x'_{n}(1 - 2{x'_{n}}^{2} + {x'_{n}}^{3})]$, $\eta = 1 - \epsilon$, and $\delta = \epsilon$. There are at least two ways in which this result undermines the claim that the demon's apprentice example demonstrates the existence of a hawkmoth effect which is an epistemological analog of the butterfly effect.

The existence of small arbitrary model perturbations demonstrated in the above theorem for first-order difference equations, of which the logistic map is an example, shows that the perturbation presented in the demon example is only one possible perturbation amongst the infinite space of admissible perturbations that are close to the logistic map under the maximum one-step error model metric. In fact, as the argument demonstrates, we can perturb our model in {\em any way} we wish and still remain as close to the initial model as desired. It should therefore be no surprise that we can find models close to the logistic map that generate trajectories in the state space vastly diverging from the logistic map over certain time intervals; indeed, we should expect to find nearby models that exhibit essentially any behavior we want, including some which vastly deviate from the logistic map across any given time interval and others which remain arbitrarily close to the logistic map for all future times. In particular, there is no a priori reason we should expect that the modified logistic map is an example of a commonly occurring small model error.  The butterfly effect is so important because numerically small differences between the true value of a system variable and its measured value are absolutely common and normal.  What reason is there to think that small model errors of the kind we would expect to find in climate science, atmospheric science, and other domains of non-linear modeling will normally produce deviations on such short timescales as they do in the demon example? Why should such a weirdly concocted example shift any burdens of proof of the kind the LSE group demand of us?   Consequently, at the very least, the demon example only retains its relevance as evidence for the epistemic force of the so-called hawkmoth effect if it is accompanied by a strong argument showing how it is precisely this sort of perturbation which is often encountered when constructing weather forecasting models. But there may be good reasons to think this is not the case (see the next point as well as Section \ref{conclusion}).

More generally, Theorem \ref{small_model_perturbations} indicates that the maximum one-step error metric is quite simply too easily satisfied and does not really get at what makes two models similar or close.\footnote{In addition to the evidence of theorem \ref{small_model_perturbations}, we also offer the following anecdotal evidence that maximum one-step error is not a particularly robust measure of model closeness.  It happens that in one of the many papers published by the LSE group on this topic, (Frigg et al., 2013(a)), they use an ever-so-slightly different version of the perturbation than they do in their other papers.  In place of the function in equation \ref{modified_logistic} they instead used 
\begin{equation}
\label{alternate_modified_logistic}
\tilde{p}_{t+1} = 4\tilde{p}_{t}(1 - \tilde{p}_{t}) \left[ (1-\epsilon) + \frac{4}{5}\epsilon( \tilde{p}_{t}^{2} - \tilde{p}_{t} + 1) \right]
\end{equation}
What is interesting is that, like in (Frigg et al, 2014), they report in this paper that for this different perturbation, at $\epsilon=.1$, the maximum one-step error (relative to the standard logistic equation) is $.005$.  But this is wrong.  The small change in the equation makes the maximum one step error skyrocket to .04, for the same value of $\epsilon$.  We can think of no better anecdotal demonstration of how artificial the maximum one-step error  is as a metric of model distance than the fact that the authors took themselves to be presenting the same perturbation twice, and it happened to differ on that metric by a factor of 10.

}  It would be difficult indeed to argue that the difference equations 
$$x_{n+1} = 2x_{n} \quad \textrm{and} \quad x_{n+1} = 2x_{n} + .004e^{x_{n}} - .03\sin x_{n}$$
are highly similar and ought to be considered ``close'' in model space simply because after one time step they do not yet produce significantly diverging trajectories in state space---the newly added sinusoidal and exponential terms behave so differently from the linear term present in the original equation that we would certainly not want to call these two models ``close". Furthermore, we would in particular definitely not expect to be able to predict well the long-term behavior of a physical system using both of these equations since the perturbations introduced in the second equation model entirely different physical dynamics.  Obviously, whether a particular metric of distance is ``good" or not, depends on what we intend to do with it.  There are lots of purposes for which the supremum metric won't be good. Two real-valued functions that are close in the supremum metric, for instance, may have derivatives that differ by arbitrarily large amounts: If one needs to draw conclusions not only about the model but also about how quickly its states change (as one often does with climate models), then the supremum metric might be a poor measure of distance. For similar reasons, two functions may be close in the supremum metric but have different numbers of maxima and minima, and their maxima and minima might be in entirely different places. If one cares about the maximal value of some model, then the supremum metric may not be appropriate. \footnote{For example, note that both of these cases apply to our example perturbation---a sine function has different numbers of local minima and maxima than a linear function, while the derivatives of the exponential and linear functions clearly differ by arbitrarily large amounts.}\textsuperscript{,}\footnote{We thank an anonymous referee for suggesting some of these arguments.}

%%%%%%%%%%%%%%%%%%%%%%%%%%%%%%%%%%%%%%
\section{Small polynomials on [0,1]}
%%%%%%%%%%%%%%%%%%%%%%%%%%%%%%%%%%%%%%

The astute reader might notice the following detail: in Frigg et al.'s demon example, they obtain a maximum one-step error between the freshman and senior apprentices of $.005$, and they do this with a value of $\epsilon$ of $0.1$.  Using the proof of Theorem \ref{small_model_perturbations}, you can calculate what value of $\epsilon$ the theorem guarantees will give you a maximum one-step that they achieve, ($.005$), and it is the relatively small number of $.0025$.  But they achieve their result with a relatively large value of $\epsilon$ of $.1$.\footnote{Note that there  is still a similarly large discrepancy even if one folds the multiplicative factor $16/5$ into the $\epsilon$ term for the purposes of comparison with Theorem \ref{small_model_perturbations}.}  What explains this?   Is the measure of model closeness reasonable if we disallow overly small values of $\epsilon$?  The very small value of epsilon that our proof produces is sufficient, but does not seem to be necessary.  Is it in fact we that are making a misleading case here?

No. Let $f(x)$ and $g(x)$ be the polynomials on $[0, 1]$ underlying (\ref{logistic2}) and (\ref{modified_logistic}), respectively. Then we have
\begin{align*}
|f(x) - g(x)| &= \left| 4x(1 - x) - \left[
(1 - \epsilon)4x(1 - x) + \frac{16\epsilon}{5}\left[ x(1 - 2x^{2} + x^{3}) \right]\right] \right| \\
&= \left| 4\epsilon x(1 - x) - {16 \epsilon \over 5}\left[ x(1 - 2x^{2} + x^{3}) \right] \right| \\
&= \left| \epsilon \left( 4 - {16 \over 5} \right)\left[ x(1 - x) - x(1- 2x^{2} + x^{3})  \right] \right| \\
&= {4\epsilon \over 5}\left| x - x^{2} - x + 2x^{3} - x^{4} \right| \\
&= {4\epsilon \over 5}\left| x^{2} - 2x^{3} + x^{4} \right|,
\end{align*}
and therefore $\sup|f(x) - g(x)| < \sup|x^{2} - 2x^{3} + x^{4}| = .0625$. This observation, that there is a polynomial with ``large" coefficients that is approximately $0$ on $[0, 1]$, in fact points to the existence of a whole space of such polynomials:

\begin{ctheorem}{2}
\label{small_polynomials}
For all $\epsilon > 0$ and $1 > \delta > 0$, there exists an infinite set of polynomials $g: [0, 1] \rightarrow [0, 1]$, written 
$$g(x) = \alpha_{n}x^{n} + \alpha_{n-1}x^{n-1} + \cdots + \alpha_{k+1}x^{k+1} + \alpha_{k}x^{k},$$
such that
\begin{equation}
\label{large_coeffs_approx}
\min\{|\alpha_{0}|, |\alpha_{1}|, \ldots, |\alpha_{n}|\} \geqslant 1 - \delta \quad \textrm{and} \quad \sup|g(x)| < \epsilon.
\end{equation}
\end{ctheorem}

\begin{figure}[H]
\begin{center}
\includegraphics[scale=0.75]{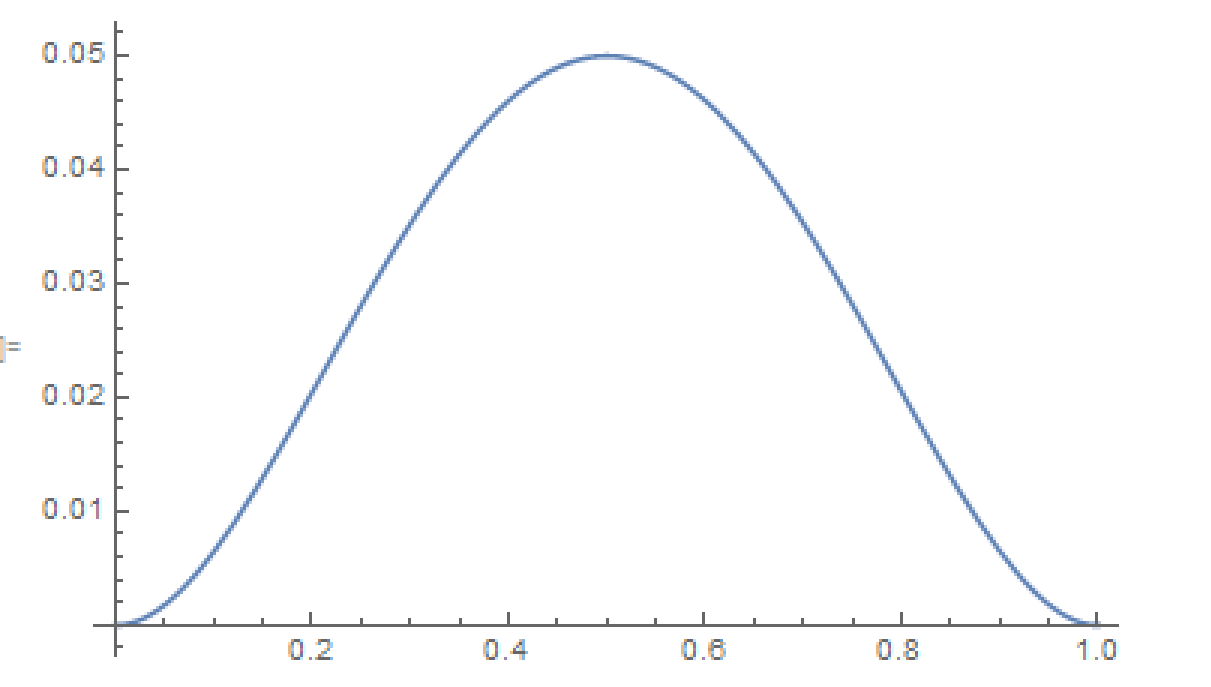}
\caption{This is a plot of the size of the freshman demon's perturbation evaluated on the interval $[0,1]$.}
\label{fig:small perturbation}
\end{center}
\end{figure}

\begin{figure}[H]
\begin{center}
\includegraphics[scale=0.75]{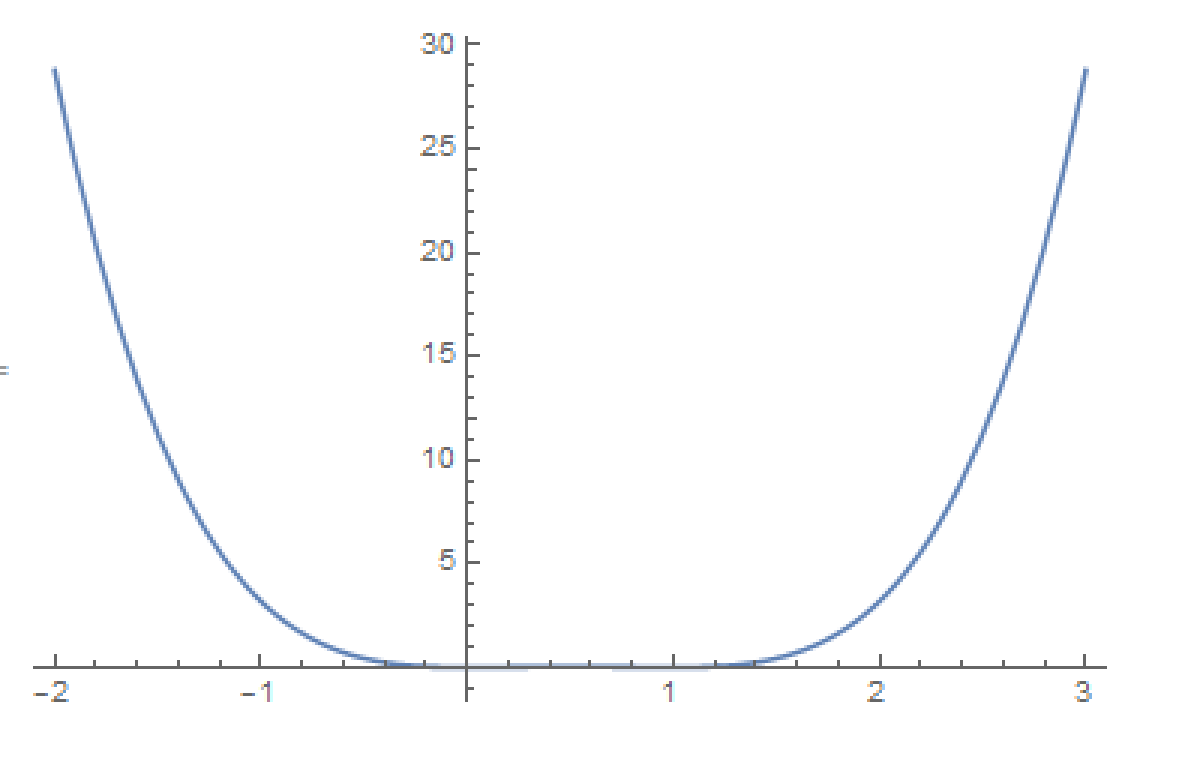}
\caption{This is a plot of the size of the freshman demon's perturbation evaluated on the interval $[-2,3]$. One can no longer even see the hump at 0.5.}
\label{fig:large perturbation}
\end{center}
\end{figure}

This explains why, in the demon case, it is possible to use a relatively large value of $\epsilon$. While Theorem \ref{small_model_perturbations} shows that maximum one-step error is a poor metric of model closeness, and in fact, provides a method of generating arbitrarily many ``close" functions on this metric, Theorem \ref{small_polynomials} shows that in certain spaces of polynomials, the Theorem \ref{small_model_perturbations} method is unnecessary to obtain ``closeness", since there are other resources for doing so. Both theorems, however, illustrate the same underlying fact: What are in fact large perturbations can be made to look very small under the right constraints.  Theorem \ref{small_polynomials}, when we are restricted to polynomial space, actually contains the more powerful method, as can be seen by looking at the two plots in Figures \ref{fig:small perturbation} and \ref{fig:large perturbation}. They clearly show that freshman demon's perturbation is not at all small, even though its maximum value on the interval [0,1] is very small.

If this is right, then it suggests that the demon example is highly atypical in its ability to exhibit what looks like fast divergence in the trajectories given small maximum one-step error and a relatively large  perturbation constant.  It is doubly atypical, in fact, in just the ways discussed above.  (It uses both the Theorem \ref{small_model_perturbations} method and the Theorem \ref{small_polynomials} method.)

%%%%%%%%%%%%%%%%%%%%%%%%%%%%%%%%%%%%%%
\section{Conclusion} \label{conclusion}
%%%%%%%%%%%%%%%%%%%%%%%%%%%%%%%%%%%%%%

In the above, we have argued that there is no hawkmoth effect that generally plagues non-linear modeling.   There is not one that accompanies the absence of structural stability, and there is not a ubiquitous feature of non-linear models that is highlighted by the demon example.  (Indeed we have seen that the focus on non-linearity by the LSE group is mysterious, since it does not figure in their arguments at all.)  But it would be wrong to interpret us as pollyannaish about climate modeling and its epistemological pitfalls.   We understand that climate modeling is difficult, and it is fraught with many possible sources of error.   Even our very best models of the climate are currently unsuitable for making some of the projections that important policy decisions might be sensitive to.  We are not unaware of this, nor do we deny it.  But the reasons for this have to do with the fact that some of the features of our climate system are poorly understood or poorly parameterized.   Regional projections are particularly difficult because of the coarse graining of the globe’s topography in global simulations, and other domain-specific features of global climate models.  None of these difficulties follow from arbitrarily small possible model errors that are magnified by a lack of structural stability.

There is a significant difference between known inadequacies in a model that are the result of idealization -- both ``dynamical" (not accounting for e.g. turbulence, the biosphere, relativistic effects, etc.) and computational (discretization, parametrization, etc.) -- and the possibility of infinitesimally small structural errors. The former is a known problem, and climate scientists and the IPCC alike are deeply concerned with eliciting the best possible estimate of the degree of uncertainty that arises from these sources. On the other hand, so far we have seen no reason to believe that the the latter, as we have demonstrated via our arguments in the previous sections, produce any significant \textit{decision-relevant} uncertainties. 

We would further add that we have no \textit{a priori} opposition to exploring the possibility of finding either precise mathematical results, or arguments by analogy to model systems, that would be suggestive of how likely we are to find something like structural chaos in a class of systems. It might very well be the case that small model errors could have outsized impacts (relative to the size of the model error) on our predictions and projections, broadly construed, and that pure math, or the investigation of toy systems, rather than empirical investigation, could guide us to this. But a research program that was serious about exploring that question would need to be much more serious about two or three questions:  

\begin{itemize}
\item  What is the universality class (or model space) for a given physical system, e.g. Earth's climate? It seems safe to at least assume that the functions defining physical models are continuous, but could we go further? Perhaps all such functions are also differentiable or even smooth (infinitely differentiable); could they also necessarily be analytic? Maybe we can even rule out specific types of functions, e.g. logarithmic or exponential functions, and argue that a given physical system must be modeled by equation(s) involving only closed-form or algebraic expressions, or something far more restrictive, such as only polynomials of degree 2 or less. At the very least, it seems potentially too bold, and certainly unsubstantiated, to assert that the universality class is as large as the space of continuous functions. If the model space is more restricted, the impact of ``small model error" could very well begin to disappear because model perturbations that do not substantially affect the predictions or projections that interest us could be far more common, perhaps even the norm.\footnote{Some of these questions are very similar to the ones posed by Mayo-Wilson in (Mayo-Wilson, 2014).}
\item What is the right metric of distance? i.e. what characteristics make one model nearby to another?  The kind of answer we give to this question might be rather different if the question is genuinely epistemological, rather than topological.

\item  If toy models are going to be used in a research program like the one that the LSE group want to conduct, what kinds of toy models are suitable? And what toy models produce idiosyncratic features like the ones we have pointed out, in sections 5 and 6, the logistic map suffers from?
\end{itemize}
But in any such research program, like any program in the philosophy of science, we believe the cardinal rule should be: do no harm.  Wildly skeptical scenarios (``poison pills,'' and the like) about a scientific program with serious policy implications should be advanced only with the greatest possible care.

In ``Probabilistic Forecasting: Why Model Imperfection Is a Poison Pill," the LSE group recommends that further research be devoted to finding an antidote for the `poison pill' (Frigg, 2013a, p. 488).  Climate models continue to be imperfect in a variety of ways that matter to policy making and decision support.  Sometimes those imperfections have radical consequences that can be intuitively characterized along the lines of structural chaos.  But we consider this paper to be an antidote for hawkmoths.

%%%%%%%%%%%%%%%%%%%%%%%%%%%%%%%%%%%%%%
\section{Appendix  A: Proof of Theorem 1}
%%%%%%%%%%%%%%%%%%%%%%%%%%%%%%%%%%%%%%

\begin{ctheorem}{1}
Suppose we are given a difference equation of the form 
\begin{equation}
\label{3}
x_{n+1} = f(x_{n}),
\end{equation}
where $x_{i} \in \mathbb{R}$ and $f: A \rightarrow B$ is an arbitrary function from the bounded interval $A \subset \mathbb{R}$ into the bounded interval $B \subset \mathbb{R}$.
Then given any function $g: A \rightarrow B$ and $\epsilon > 0$, there exists $\delta > 0$ and $\eta > 0$ such that the maximum one-step error of
\begin{equation}
\label{13}
x'_{n+1} = \eta f(x'_{n}) + \delta g(x'_{n}),
\end{equation}
from (\ref{3}) is at most $\epsilon$ and $x'_{n+1} \in B$.
\end{ctheorem}
\begin{proof}
Set $\delta$ and $\eta$ such that 
$$|\delta| \leqslant \left|\frac{\epsilon}{2\sup\{g(x_{n})\}}\right| \quad \textrm{and} \quad |\eta-1| \leqslant \left|\frac{\epsilon}{2\sup\{f(x_{n})\}}\right|,$$
where $\sup\{f(x)\}$ denotes the supremum\footnote{Essentially, for sufficiently ``well-behaved" functions, the maximum value.} of $f$ over all $x \in \mathbb{R}$. Note that the suprema exist\footnote{That is, the suprema are finite.} because $f$ and $g$ are bounded. Since the one-step error is
\begin{align*}
|x'_{n+1} - x_{n+1}| &= |\eta f(x_{n}) + \delta g(x_{n}) - f(x_{n})| \\
&= |(\eta - 1)f(x_{n}) + \delta g(x_{n})|,
\end{align*}
by the triangle inequality, the maximum one-step error is 
\begin{align*}
\sup\{|x'_{n+1} - x_{n+1}|\} &= \sup\{|(\eta - 1)f(x_{n}) + \delta g(x_{n})|\} \\
&\leqslant \sup\{|\eta - 1||f(x_{n})| + |\delta||g(x_{n})|\} \\
&\leqslant \sup\left\{\left|\frac{\epsilon}{2\sup\{f(x_{n})\}}\right||f(x_{n})|\right\} + \sup\left\{\left|\frac{\epsilon}{2\sup\{g(x_{n})\}}\right||g(x_{n})|\right\} \\
&= \left|\frac{\epsilon\sup\{f(x_{n})\}}{2\sup\{f(x_{n})\}}\right| + \left|\frac{\epsilon \sup\{g(x_{n})\}}{2\sup\{g(x_{n})\}}\right| \\
&= \frac{\epsilon}{2} + \frac{\epsilon}{2} \\
&= \epsilon,
\end{align*}
as desired.
\end{proof}

%%%%%%%%%%%%%%%%%%%%%%%%%%%%%%%%%%%%%%
\section{Appendix  B: Proof of Theorem 2}
%%%%%%%%%%%%%%%%%%%%%%%%%%%%%%%%%%%%%%

\begin{ctheorem}{2}
\label{small_polynomials}
For all $\epsilon > 0$ and $1 > \delta > 0$, there exists an infinite set of polynomials $g: [0, 1] \rightarrow [0, 1]$, written 
$$g(x) = \alpha_{n}x^{n} + \alpha_{n-1}x^{n-1} + \cdots + \alpha_{k+1}x^{k+1} + \alpha_{k}x^{k},$$
such that
\begin{equation}
\label{large_coeffs_approx}
\min\{|\alpha_{0}|, |\alpha_{1}|, \ldots, |\alpha_{n}|\} \geqslant 1 - \delta \quad \textrm{and} \quad \sup|g(x)| < \epsilon.
\end{equation}
\end{ctheorem}
\begin{proof}
Let $0 \leqslant e^{2}(n-k)/\epsilon < k < n \leqslant 1/\delta$, where $e = 2.71828\ldots$ is Euler's constant, and define $\alpha_{k}, \ldots, \alpha_{n}$ by setting
\begin{align}
\label{small_poly_conditions}
\sum_{i=k}^{n}\alpha_{i} = 0, \quad |1 - |a_{i}|| < \delta, \quad \textrm{and} \quad |\alpha_{i+1} + \alpha_{i}| < {\epsilon \over n-k}
\end{align}
for all $k \leqslant i \leqslant n$. Then since
\begin{align*}
\sup|g(x)| &= \sup|\alpha_{n}x^{n} + \alpha_{n-1}x^{n-1} + \cdots + \alpha_{k+1}x^{k+1} + \alpha_{k}x^{k}| \\
&\leqslant \sup|\alpha_{n}x^{n} + \alpha_{n-1}x^{n-1}| + \cdots + \sup|\alpha_{k+1}x^{k+1} + \alpha_{k}x^{k}|,
\end{align*}
it suffices to determine the extrema of $g_{i}(x) := \alpha_{i+1}x^{i+1} + \alpha_{i}x^{i}$ for all $k \leqslant i \leqslant n-1$. In that direction, note that the extrema of a function occurs either where that function's first derivative vanishes or at the end points. Thus, the possible maxima are $g_{i}(0) = 0$, 
$$|g_{i}(1)| = |\alpha_{i+1} + \alpha_{i}| < {\epsilon \over n-k},$$
by (\ref{small_poly_conditions}), and, since
\begin{align*}
0 = {dg_{i}(x) \over dx} &= x^{i}(\alpha_{i+1}x + \alpha_{i}) \\
&= ix^{i}(\alpha_{i+1}x + \alpha_{i}) + \alpha_{i+1}x^{i} \\
&= i\alpha_{i+1}x^{i+1} + (i\alpha_{i} + \alpha_{i+1})x^{i} \\
&= x^{i}(i\alpha_{i+1}x + i\alpha_{i} + \alpha_{i+1})
\end{align*}
has solutions at $x = 0$ and $x = -(i\alpha_{i} + \alpha_{i+1})/i\alpha_{i+1}$, 
\begin{align*}
\left| g_{i}\left(-{i\alpha_{i} + \alpha_{i+1}\over i\alpha_{i+1}}\right) \right| &= \left|\alpha_{i+1}\left( -{i\alpha_{i} + \alpha_{i+1}\over i\alpha_{i+1}} \right)^{i+1} + \alpha_{i}\left( -{i\alpha_{i} + \alpha_{i+1}\over i\alpha_{i+1}} \right)^{i} \right| \\
&= \left|(-1)^{i+1}{\alpha_{i+1}(i\alpha_{i} + \alpha_{i+1})^{i+1}\over (i\alpha_{i+1})^{i+1}} + (-1)^{i}{\alpha_{i}(i\alpha_{i} + \alpha_{i+1})^{i}\over (i\alpha_{i+1})^{i}} \right| \\
&= \left|{\alpha_{i+1}(i\alpha_{i} + \alpha_{i+1})^{i+1} - i\alpha_{i}\alpha_{i+1}(i\alpha_{i} + \alpha_{i+1})^{i}\over (i\alpha_{i+1})^{i+1}} \right| \\
&= \left| {\alpha_{i+1}^{2}(i\alpha_{i} + \alpha_{i+1})^{i}\over (i\alpha_{i+1})^{i+1}} \right|.
\end{align*}
Applying (\ref{small_poly_conditions}), the definition of $k$ and $n$, and the well-known inequality $(1 + 1/i)^{i} \leqslant e$, we have
$$\left| g_{i}\left(-{i\alpha_{i} + \alpha_{i+1}\over i\alpha_{i+1}}\right) \right| < {i^{i}(1 + \delta)^{i} \over i^{i+1}\alpha_{i+1}^{i-1}} = {(1 + \delta)^{i} \over i(1 - \delta)^{i-1}} \leqslant {(1 + 1/i)^{i} \over i(1 - 1/i)^{i-1}} \leqslant {e^{2} \over k} < {\epsilon \over n-k}.$$
Hence, after summing over all $i$, we conclude that $\sup|g(x)| < \epsilon$. Since the first inequality in (\ref{large_coeffs_approx}) holds by definition, and there exist an uncountable infinity of coefficients $\alpha_{k}, \ldots, \alpha_{n}$ satisfying (\ref{large_coeffs_approx}), this yields the desired result. 
\end{proof}

\end{document}